\documentstyle{mn}

\title[TW Pic]
      {Is TW Pictoris really an intermediate polar?}

\author[Norton \& Beardmore]
       {A.J. Norton$^{1}$ and A.P. Beardmore$^{2}$\\
        $^1$Department of Physics, The Open University, Walton Hall, 
        Milton Keynes MK7 6AA \\
	$^2$Department of Physics, The University of Keele, Keele,
	Staffordshire ST5 5BG}

\date{Accepted 199? Month ??.
      Received 199? Month ??; 
      in original form 199? Month ??}

\begin{document}

\maketitle

\begin{abstract}

We present the results of a long {\em ROSAT} HRI observation of the 
candidate intermediate polar TW Pic. The power spectrum shows no sign
of either the previously proposed white dwarf spin period or the proposed 
binary orbital period (1.996~hr and 6.06~hr respectively). The limits to 
the X-ray modulation are less than 0.3\% in each case. In the absence of a 
coherent X-ray pulsation, the credentials of TW Pic for membership of the 
intermediate polar subclass must be suspect. We further suggest that the 
true orbital period of the binary may be the shorter of the two previously 
suggested, and that the longer period may represent a quasi-periodic 
phenomenon associated with the accretion disc.

\end{abstract}


\section{INTRODUCTION}

TW Pic is a 14th magnitude cataclysmic variable which was identified
by Tuohy et al (1986) as the optical counterpart to the {\em HEAO-1} X-ray 
source H0534--581. Although Tuohy et al (1986) failed to detect any
periodicities in either optical photometry or {\em EXOSAT} X-ray observations, 
they suggested that TW Pic may be an intermediate polar based on its X-ray 
to optical flux ratio and its strong He{\sc ii} 4686\AA \ emission line
(for a review of intermediate polars, see Patterson 1994). Subsequently, 
Buckley \& Tuohy (1990) reported an optical 
spectroscopic study of the system and determined periods of $2.1 \pm 0.1$~hr 
and $6.5 \pm 1.0$~hr from radial velocity measurements, which they took to 
represent the spin period of the white dwarf and the orbital period of the 
binary, respectively. A re-analysis of the {\em EXOSAT} data by Buckley \&
Tuohy (1990) this time revealed the 2.1~hr proposed spin period, and new 
optical photometry also showed the 6.5~hr proposed orbital period. 
A later optical photometric study of TW Pic by Patterson \& Moulden (1993)
revealed periods of $6.06 \pm 0.03$~hr and $1.996 \pm 0.001$~hr which they
interpreted as more accurate determinations of the orbital and spin
periods found by Buckley \& Tuohy (1990). They noted, however, that the 
6.06~hr period appeared to drift throughout their 3 weeks of observations,
and that the supposed orbital profile varied greatly from cycle to cycle.

\section{A {\em ROSAT} HRI OBSERVATION OF TW PIC} 

Here we report details of a long X-ray observation of TW Pic,
made with the {\em ROSAT} High Resolution Imager (Zombeck et al 1995). 
The observation comprises 47.9~ksec on source between 1995 Nov 15 17:39 UT and 
1995 Nov 24 19:59 UT, and was performed in `time critical' mode, 
observing the source for about five hours per day, over ten days.
Using the Starlink {\em Asterix} software (Allen \& Vallance 1995), data 
were optimally extracted from a region about an arc min in radius, centred 
on the source, and background subtraction was carried out using the data 
from a concentric annulus. A binned light curve with 10~s time resolution
was created, and a lower resolution version of this is shown in Figure 1.
Each block of data represents one satellite orbit, and each panel contains
one day of observations. The mean count rate is 0.20~c~s$^{-1}$. 

\begin{figure*}
\setlength{\unitlength}{1cm}
\begin{picture}(10,18.5)
\put(-1,0){\includegraphics{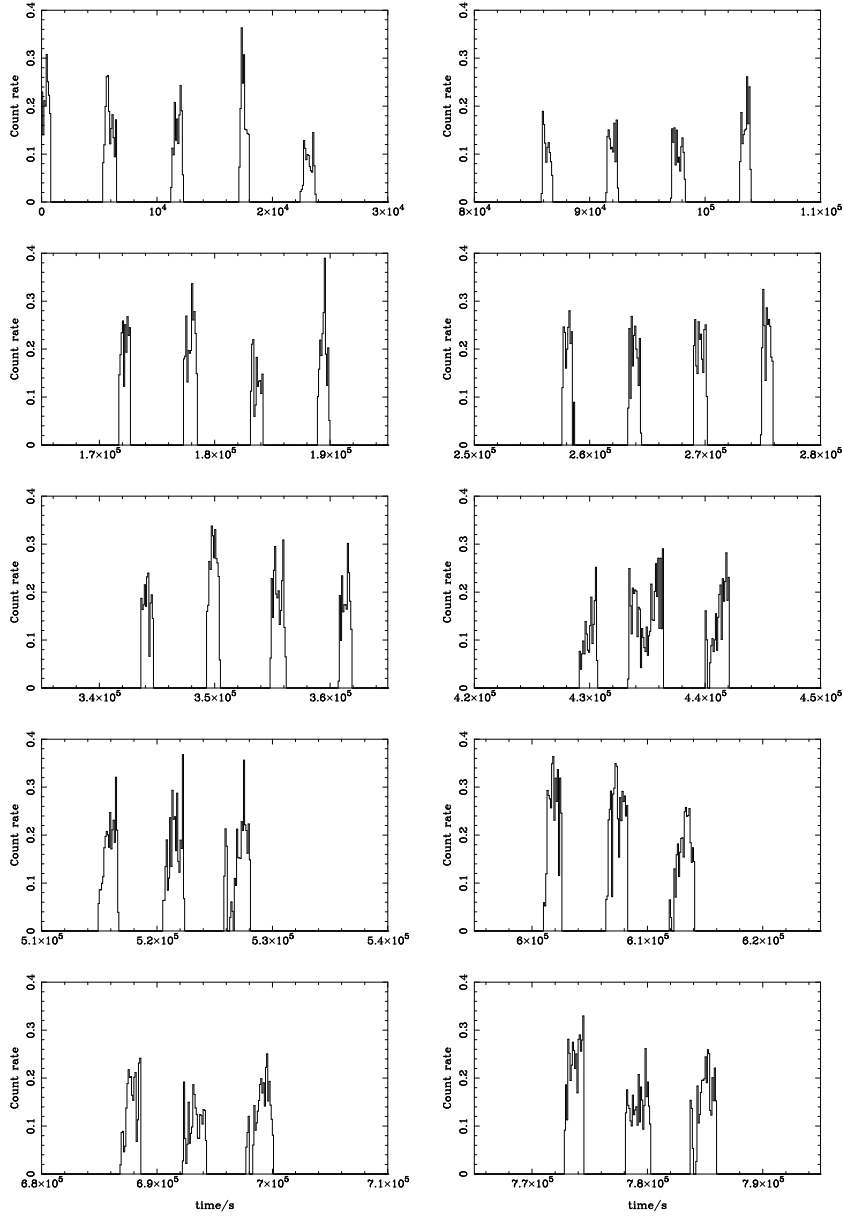}}
\end{picture}
\caption{The {\em ROSAT} HRI lightcurve of TW Pic at a time resolution
of 100~s. Each panel represents one day of observations and contains between
three and five satellite orbits of data.}
\end{figure*}            

To analyse the light-curve we used the 1-dimensional {\sc clean} algorithm
in the implementation of H.J. Lehto. This period searching algorithm is
ideal for searching for multiply periodic signals in unevenly spaced data 
and we have shown in the past that it is particularly suited to analysing 
X-ray light curves of intermediate polars (see for example Norton et al 1997, 
Beardmore et al 1998, and references in both). The results of {\sc clean}
are shown in Figure 2. 

\begin{figure*}
\setlength{\unitlength}{1cm}
\begin{picture}(10,18.5)
\put(-1,0){\includegraphics{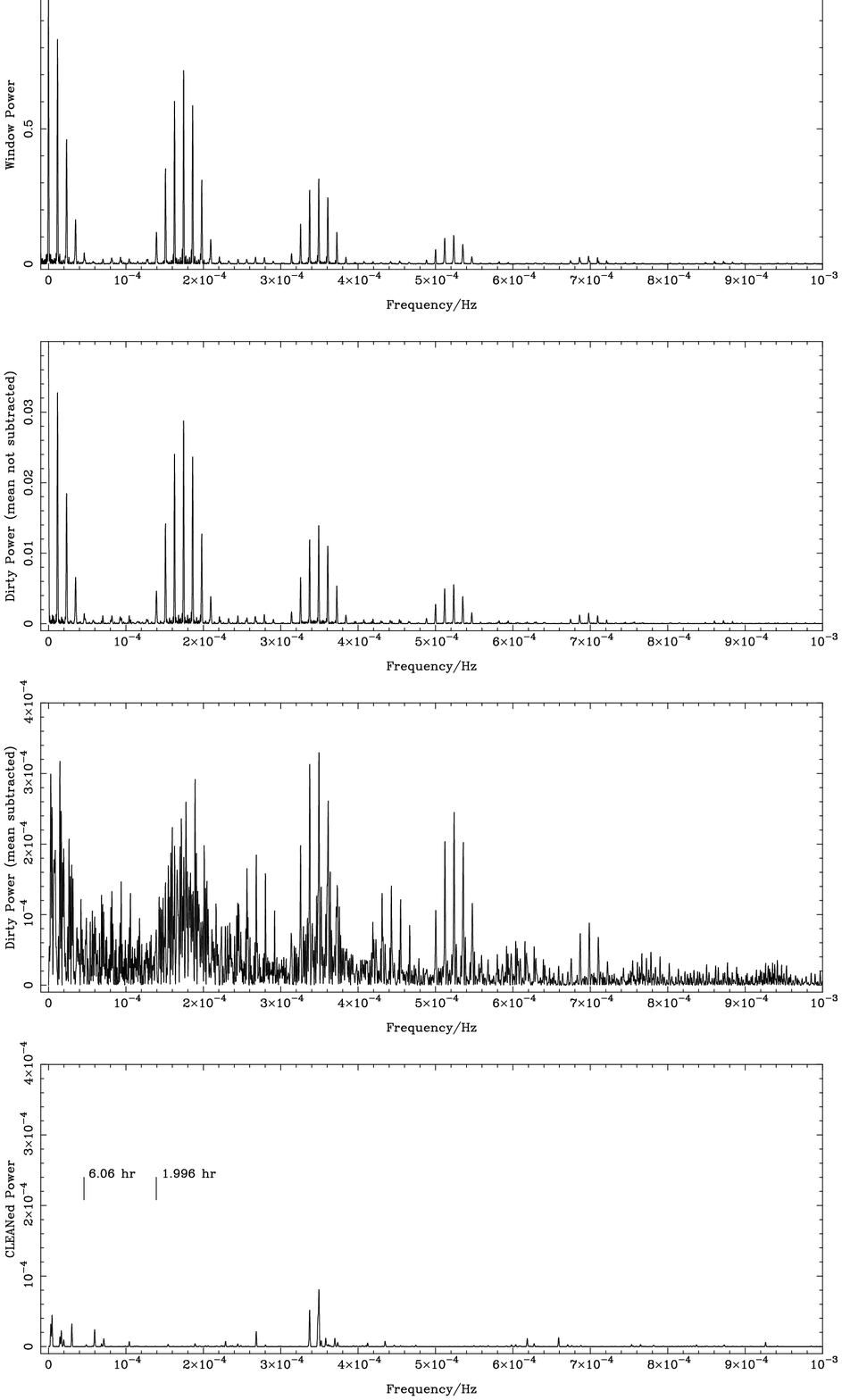}}
\end{picture}
\caption{The power spectrum of the {\em ROSAT} HRI lightcurve of TW Pic.
The top panel shows the window function of the data; the second panel
shows the `dirty' power spectrum; the third panel shows the `dirty' power 
spectrum after the mean has been subtracted from the data (note the scale 
change by a factor of 100 between the second and third panels); and the bottom 
panel shows the {\sc clean}ed power spectrum. Tick marks show the locations 
of the previously reported periods for this system.}   
\end{figure*}            

The top panel simply shows the window function of the time series -- it 
reflects the 96~min satellite orbit (large scale `envelope' structure) and 
the daily observation slots (small scale `spike' structure). The second
panel is the raw or `dirty' power spectrum of the time series. The fact that
this looks essentially the same as the window function is a good indication
that there are no periodicities in the data which result from TW~Pic itself.
The third panel is another `dirty' power spectrum of the time series, but
this time with the mean value of the data removed prior to calculation of
the Fourier transform. This effectively removes the `first order' window
function from the data allowing any weak signals to be seen more clearly. 
Note that the vertical scale of this power spectrum is 100 times greater 
than in the second panel. Some residual structure appears
to be present in this power spectrum, but close inspection again reveals
that all the peaks are at window function frequencies.

The bottom panel shows the {\sc clean}ed power spectrum, with the same 
vertical scale as the third panel. The two largest spikes in the 
{\sc clean}ed power spectrum, near to $3.5 \times 10^{-4}$~Hz, are at window 
function frequencies and so are unlikely to represent real signals. 
(In fact, their frequencies correspond to {\em half} the orbital period of 
the satellite and its one day alias.) The {\sc clean}ed X-ray power spectra 
of confirmed intermediate polars generally show clear signals at the system 
periods (see for example Norton et al 1997, Beardmore et al 1998),
and that is not the case here. There are no significant signals at either of 
the previously reported periods of this object. At periods of both 6.06~hr and 
1.996~hr, the power is less than about  $10^{-7}$~c$^{2}$~s$^{-2}$, 
corresponding to a limiting amplitude in the 
light curve of $<6 \times 10^{-4}$~c~s$^{-1}$. (Nb. The amplitude is equal 
to twice the square root of the {\sc clean}ed power.) The upper limit to any 
modulation in the X-ray light curve at these periods is therefore 0.3\%. 
Furthermore, there are no other significant periods detected either -- the
Nyquist frequency of the time series is around $3 \times 10^{-3}$~Hz and 
there are no significant signals detected out to this frequency. (The power
spectra in Figure 2 are only plotted out to $10^{-3}$~Hz to emphasize
the structure at lower frequencies.) We therefore conclude that the light 
curve of TW~Pic from this {\em ROSAT} HRI observation displays no intrinsic 
periodic signals, with a period greater than about 300~s, down to a limiting 
modulation of less than 1\%.

\section{DISCUSSION}

The {\em ROSAT} HRI count rate from TW Pic (0.20~c~s$^{-1}$) is comparable 
to that of confirmed intermediate polars with similar optical fluxes. For 
instance AO~Psc and V1223~Sgr are both 13th magnitude intermediate polars and
exhibit HRI count rates of 0.22~c~s$^{-1}$ and 0.41~c~s$^{-1}$ respectively
(Taylor et al 1997), whilst the 15th magnitude TX~Col has an HRI count rate
of 0.07~c~s$^{-1}$ (Norton et al 1997). On the basis of its X-ray to optical 
flux ratio therefore, TW~Pic would seem to be a good candidate for 
intermediate polar status, as Tuohy et al (1986) originally suggested.
However, the one unambiguous signature of an intermediate polar is a 
coherent X-ray pulsation at a period significantly less than the binary 
orbital period. Based on this most sensitive X-ray observation yet
of TW~Pic, no such pulsation exists and so we have no evidence
to support the proposed classification of the system. TW~Pic {\em may} still
be an intermediate polar, but one seen at a relatively low inclination angle
such that a roughly constant X-ray flux is seen from the upper magnetic
pole, and the lower one is permanently hidden. In this case, we would not 
expect the spin period of the white dwarf to be apparent in optical photometry
or spectroscopy either. So, whether or not TW~Pic is an intermediate polar,
we must therefore suggest some other explanation for the periods previously 
reported in this system.

The $\sim 2$~hr period cannot represent the spin period of an accreting 
magnetic white dwarf, otherwise we would have seen evidence for it in our 
{\em ROSAT} HRI data. The original detection of this period was in the optical 
spectroscopic data of Buckley \& Tuohy (1990). However, their subsequent 
detection of a similar period in the {\sl EXOSAT} X-ray data is barely 
significant, and indeed went un-noticed in their earlier analysis of 
the same data (Tuohy et al 1986). It is doubtful whether the X-ray detection 
of the period would have been claimed without the prior discovery of the 
$\sim 2$~hr optical period. Nonetheless, a $\sim 2$~hr period has been 
clearly detected in both optical spectroscopy (Buckley \& Tuohy 1990) and in 
optical photometry (Patterson \& Moulden 1993), and is undoubtedly real. We 
suggest, therefore, that 1.996~hr represents the {\em orbital} period of the 
system and as such we would not necessarily expect to detect it in an 
X-ray observation, unless the system were at a relatively high inclination. 

The reasons for disbelieving that the $\sim 6$~hr period represents
the orbital motion have already been discussed by Patterson \& Moulden (1993).
In their photometric data, this period was observed to drift over the
course of 3 weeks by up to 0.3 cycles. Moreover, they note that the profile 
of the $\sim 6$~hr modulation simply does not look stable, and varies 
significantly from cycle to cycle. The spectroscopic and photometric 
observations reported by Buckley \& Tuohy (1990) both detected a $\sim 6$~hr 
period, but again the constancy of the period is not convincing, and in their 
case {\em could} merely be a second harmonic of the $\sim 2$~hr period, as
neither of the periods are determined very accurately. We suggest 
instead that this quasi-periodic 6~hr modulation may arise from a phenomenon 
associated with the accretion disc.

A comparable system may be the intermediate polar TV~Col (Hutchings et al 
1981), which exhibits a spectroscopic (orbital) period of 5.49~hr but a 
photometric period of 5.19~hr (in addition to the white dwarf spin period of 
1911~s). The additional presence of a $\sim 4$~d `beat' period (such that 
$1/5.49~{\rm hr} + 1/4~{\rm d} = 1/5.19~{\rm hr}$) led Barrett, O'Donoghue 
and Warner (1988) to interpret these multiple periodicities in terms of a 
retrogradely precessing accretion disc. The disc is assumed to precess with a 
4~d period, and the 5.19~hr period may then arise due to tidal interactions 
with the secondary (Hellier 1993). Augusteijn et al (1994) noted that both 
the 5.19~hr and 4~d periods in TV~Col are unstable and may vary monatonically. 
Given the similarly unstable $\sim 6$~hr period in TW~Pic, disc precession may 
be the explanation for the long period seen in this system also.

Finally, we note that if TW~Pic really were a magnetic system, and if the spin 
period of the white dwarf really were around two hours, it would represent 
the {\em slowest} rotator amongst all the intermediate polars. The only 
confirmed intermediate polar coming close to this is EX~Hya, with a spin 
period of 67~min. However, EX~Hya is unusual in that its spin and orbital 
periods are in a $\sim$2:3 ratio. G. Wynn (private communication) has shown 
that high magnetic field intermediate polars will indeed evolve such that 
their periods end up in a $\sim$2:3 ratio, not the apparent $\sim$1:3 ratio 
which the previously proposed periods of TW Pic might indicate.

In conclusion, we find that there is no convincing evidence for regarding  
TW~Pic as an intermediate polar, and we suggest that the true orbital period
of the binary is the shorter of the two previously identified periods.

\section*{ACKNOWLEDGMENTS}

The data analysis reported here was carried out using facilities provided 
by PPARC, Starlink and the Open University Research Committee.

\bsp

\end{document}